\documentclass[aps,pre,onecolumn,superscriptaddress,showpacs,floatfix]{revtex4}
\usepackage{graphicx}
\usepackage{epsfig}

\newcommand{\bs}{\symbol{92}}

\begin{document}

\title{Nanomorphology of the blue iridescent wings of a giant tropical wasp,\\
\textit{Megascolia procer javanensis} (Hymenoptera)}

\author{Micha\"{e}l Sarrazin}
\email{michael.sarrazin@fundp.ac.be}\affiliation{Laboratoire de
Physique du Solide, Facult\'es Universitaires Notre-Dame de la
Paix,\\rue de Bruxelles, 61, B-5000 Namur Belgium}

\author{Jean Pol Vigneron}
\affiliation{Laboratoire de Physique du Solide, Facult\'es
Universitaires Notre-Dame de la Paix,\\rue de Bruxelles, 61,
B-5000 Namur Belgium}

\author{Victoria Welch}
\affiliation{Laboratoire de Physique du Solide, Facult\'es
Universitaires Notre-Dame de la Paix,\\rue de Bruxelles, 61,
B-5000 Namur Belgium}

\author{Marie Rassart}
\affiliation{Laboratoire de Physique du Solide, Facult\'es
Universitaires Notre-Dame de la Paix,\\rue de Bruxelles, 61,
B-5000 Namur Belgium}

\date{\today}

\begin{abstract}
The wings of the giant wasp \textit{Megascolia Procer Javanensis}
are opaque and iridescent. The origin of the blue-green
iridescence is studied in detail, using reflection spectroscopy,
scanning electron microscopy and physical modelling. It is shown
that the structure responsible for the iridescence is a single
homogeneous transparent chitin layer covering the whole surface of
each wing. The opacity is essentially due to the presence of
melanin in the stratified medium which forms the mechanical core
of the wing.
\end{abstract}


\maketitle





%
%
%




\section{Introduction}

The structural coloration of living organisms is currently
receiving much attention \cite
{zi-nas-03,Vigneron-pre-2006-magpie,Parker-nat-2001,welch-pre-2006,Parker-nat-2003,Ghiradella-sci-1972,Tada-ao-1998,Yoshioka-prsl-2004,Lawrence-ao-2002,Vukusic-ao-2002,Argyros-mi-2002,Vertezy-ma-2004}%
. The keys to the visual effects occurring in insects -- mainly
butterflies and beetles -- are being progressively revealed, and
the relationship between the cuticle's nanomorphology and its
optical properties is becoming ever more accurate, often with the
support of physical modelling and numerical simulations. In a few
cases, artificial structures which mimic the natural functions
\cite{Vigneron-pre-vit-06} have consolidated our understanding of
the colouration mechanisms.

However, some families of insects, though visually appealing, have
been studied relatively less. In this work, we report on the
analysis of the iridescence of the wings of a giant wasp,
\textit{Megascolia procer javanensis} (Betrem \& Bradley 1964).
Wasps are winged insects which belong to the order Hymenoptera,
which also includes ants and bees. Although some wasps have
developed social behaviors, the vast majority of the 200,000
species are solitary. Wasps can be found in most regions of both
hemispheres. Some are solid black or dark blue, but most of them
display conspicuous red, orange, or yellow markings. The wings are
opaque or transparent.

\begin{figure}[t]
\centerline{\ \includegraphics[width=8.0 cm]{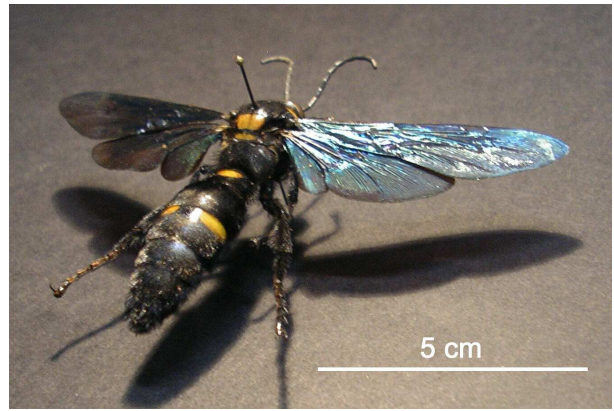}}
\caption{(Colour online) A collected specimen of male
\textit{Megascolia procer javanensis} (Hymenoptera). Note the dark
wings of this Scoliid, showing bluish reflections.} \label{fig1}
\end{figure}

\begin{figure}[b]
\centerline{\ \includegraphics[width=8.0 cm]{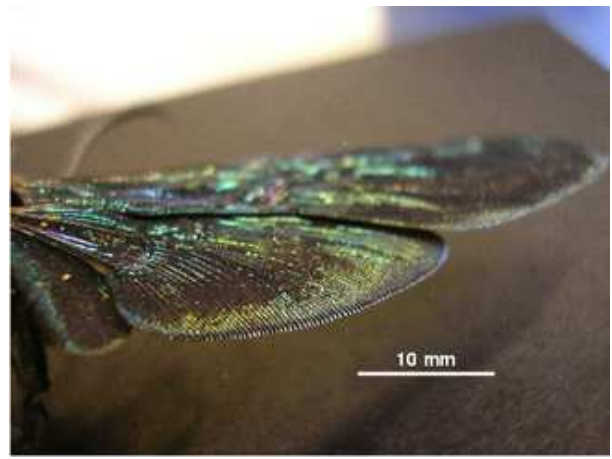}}
\caption{(Colour online) The wings of \textit{Megascolia procer}
are highly sophisticated organs, balancing low inertia, strength
and optimized aerodynamics. Note the rippled surface, which
produce a wavy cross-section for the bearing membrane. }
\label{fig2}
\end{figure}

\textit{Megascolia procer javanensis} is a large and robust
insect, about 5 cm in length, which belongs to the small family of
Scoliidae\cite {Osten-lbb-2000} (see Fig. \ref{fig1}). Organisms
in this family have long been observed and studied in relation to
their parasitic behaviour, in particular by the French naturalist
Jean-Henry Fabre\cite{Fabre-se-1891}. Members of these Scoliidae
are indeed external parasites of Scarabaeid
larvae, which means that they are able to sting and thereby paralyze a grub%
\cite{Vereecken-nfg-2003}, lay an egg on it, and leave it in the
soil, so that the developing larva will feed on the grub. The
specimen under study here \cite{Betrem-zm-1964-1,Betrem-zm-1964-2}
originates from the island of Java. The body of this insect is
slightly hairy and the wings show a large number of parallel,
longitudinal, wrinkles (see Fig. \ref{fig2}). Moreover, in this
particular species, the wings appear black and mostly opaque, with
iridescent green to bluish green reflections visible at increasing
viewing angles.

The objective of the present paper is to clarify the relationship
between the physical structure of the wings, as revealed by
scanning electronic microscopy (SEM), and their optical
properties. The next sections will show that the wings can be
modelled by a thin optical layer probably made of chitin covering
a simple chitin/melanin mixture substrate. In order to confirm
this interpretation, the experimental spectra of the scattered
light will be compared with the results of numerical simulations
based on the thin layer model.

\section{Nanomorphology}

\begin{figure}[t]
\centerline{\ \includegraphics[width=8.0 cm]{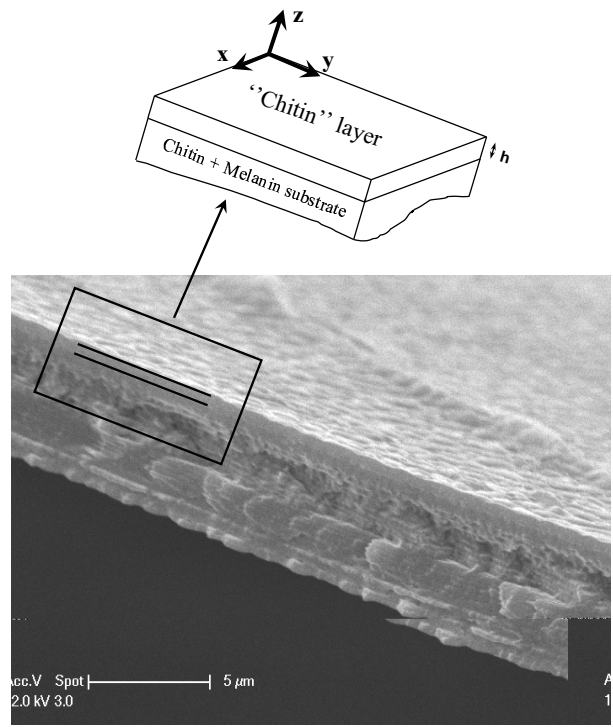}}
\caption{\textit{Megascolia Procer Javanensis} wing section
(scanning electronic microscopy picture) and model used for
simulations.} \label{fig3}
\end{figure}

Wasp wings are made from a material containing chitin, proteins
and melanin, just like the insect's cuticle. A scanning electron
microscope image of a wing, fractured in the direction normal to
its surface, reveals that the wings are covered by a thin layer of
unknown medium, with a measured average thickness $300$~nm (see
Fig. \ref{fig3}). A similar uniform layer coverage has also been
observed in other insects, such as dragonflies \cite
{Hooper-oe-2006}. The bulk of the wing, below the thin layer, is
structured as a multilayer, likely to improve mechanical strength.
The thickness of the
layers varies along the length of the wing (from about $400$~nm to $1$~$\mu $%
m), plausibly to provide a variable flexibility. Except for its
melanin content, which leads to opacity, this multilayer is
probably not directly involved in any blue-green colouring
process, because it has a thickness over $400$~nm, with an average
refractive index above $1.5$, making it a
Bragg mirror that essentially reflects at wavelengths longer than about $%
\lambda ~=~1200$~nm: well in the infrared. An harmonic of such a
resonance could, in principle, be found in the visible, near $600$
nm (orange-red), but this is actually not observed. These findings
imply implies that the multilayer is too lightly contrasted and/or
is too absorbent to produce multiple interface scattering.

Another reason to rule out the bulk wing structure as a possible
origin of the iridescence is that the layer thickness varies along
the wing. If such a structure were to be selectively reflecting,
its central colour would vary drastically along the length of the
wing and this, again, is not observed. The blue-green hue of the
reflection is very uniform on the forewing and hindwing surfaces.

As Fig. \ref{fig3} shows, the surface of the wing is slightly
corrugated, with randomly distributed rounded protrusions. A
typical distance between the protrusions' centres is 1.2~$\mu$m,
which is also their diameter. The protrusions then form a
disordered field of touching islands, with an overall thickness of
about $80$~nm. This structure, again, should not seriously impact
upon the colour production, but could be expected to broaden the
reflection both in the spectral- and emergence angles- domains.

\section{Optical properties of the wing membrane}

The reflection factor of the wing membrane was measured for
several incidence angles, in the specular geometry. For this
purpose, a piece of wing was cut from a dry specimen and glued on
a black substrate. The reflection spectra were obtained using an
Avaspec 2048/2 fibre optic spectrophotometer and the reflected
light was compared with a diffuse PTFE-based white reference tile.
This normalization produces the \symbol{92} reflection factor''
shown in Fig. \ref{fig4}. This quantity is closely related to the
reflectance, which expresses the reflected power in units of the
incident power. In the reflectance range of interest, due to the
flat response of the white standard, these quantities differ only
by a normalization factor.

The results of the measurements are given in Fig. \ref{fig4},
where the red curves (experimental results) describe the spectral
response of a wing under varying incidences. All theses
measurements were performed on a forewing, in a specular geometry
(with an emergence angle equal to the incidence angle, both
measured from the normal to the wing surface). The incidence plane
was directed along the length of the wing.

At normal incidence the backscattering measurement reveals
oscillations with reflection maxima for wavelengths near 325~nm,
505~nm and 1015~nm. The broad lineshape of these reflection band
is reminiscent of the single slab resonances, except for the
increase of the reflection maxima. The overall effect obtained is
a lack of reflected intensities between 600 and 800~nm, which
basically covers the orange-red colorimetric region. This is
consistent with the blue-green colouration of the wing. Smooth
oscillations of the reflectance spectrum, as a function of the
wavelength, indicate the interference of a small number of waves
and this is consistent with the interference occuring in the thin
layer and with a low refractive index contrast at the substrate
interface.

When the angle of incidence is increased, the spectrum is slightly
blue-shifted, as one would expect from this single slab
interference mechanism. It is important to know the refractive
indices in order to make more precise predictions and to build an
accurate model of the wing's optical behaviour. This is the
subject of the next section.

\section{Detailed modelling}

\label{model}

Electron microscopy reveals a wing made of a stack of \bs
mechanical'' (thin and rigid) slabs covered by a biopolymer layer.
The thickness of the mechanical layers varies along the wing,
while the thickness of the upper surface layer is constant. On the
other hand, the hue selection is constant over the whole surface
of the wings, which suggests that the stack multilayer should be
considered an homogeneous and very absorbant substrate and the
upper surface layer (of thickness about $300$~nm), the optical
filter. The upper layer is transparent, and its dielectric
function is likely to be close to that of chitin (indeed, in
ethanol, with a refractive index $1.4$, close enough to that of
chitin, the iridescence of the wasp's wing is reversibly
suppressed, as Fig. \ref{fig5} shows).

\begin{figure}[t]
\centerline{\ \includegraphics[width=5.5 cm]{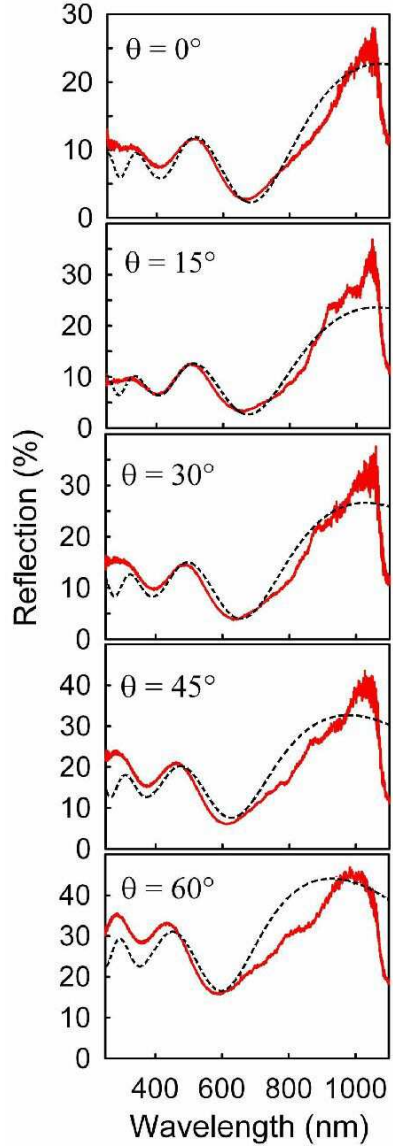}}
\caption{(Colour online) Wings reflection as a function of the
wavelength for various incidence angles. Experimental data (solid
line) and simulations results (dashed line). } \label{fig4}
\end{figure}

In the context of a single optical slab model, the reflection coefficient $%
\mathcal{R}$ is given by

\begin{equation}
\mathcal{R}=\left| \frac{r_1+r_2e^{2i\beta }}{1+r_1r_2e^{2i\beta
}}\right| ^2 \label{1}
\end{equation}
where $r_1$ ($r_2$) is the Fresnel reflection coefficient on the
air/slab interface (on the slab/substrate interface), and
\begin{equation}
\beta =\frac{2\pi }\lambda h{\sqrt{n_L^2-\sin ^2i}}  \label{2}
\end{equation}
$h$ is the slab thickness, $n_L$ the refraction index of the upper layer, $%
\lambda $ the wavelength of the incident light and $i$ the angle
of incidence. In this context, the condition for constructive
interference in the reflected beam under the incidence $i$ is that
the incident wavelength verifies (see Fig. \ref{fig6})
\begin{equation}
\lambda _{\max }=\frac{{2h\sqrt{n_L^2-\sin ^2i}}}m  \label{3}
\end{equation}
where $m$ is a positive integer. By contrast, the condition for
destructive interference in the reflected beam is that the
incident wavelength verifies
\begin{equation}
\lambda _{\min }=\frac{{4h\sqrt{n_L^2-\sin ^2i}}}{2m+1}  \label{4}
\end{equation}
The wavelengths related to the main reflection maxima or minima
for many incidence $i$ are known from experiment. Using Eqs. 3 and
4 it is then possible to fit both the overlayer thickness $h$ and
its refractive index $n_L$. The precise dispersion of the upper
layer material remains unknown and it can be neglected in the
present case. One gets $h=286$ nm (that is in agreement with SEM
assessment) and $n_L=1.76$ (in accordance with the above
hypothesis). Though the basic molecule of chitin is well defined -
(C$_8$H$_{13}$O$_5$N)$_n$ - the full composition of the hard
surfaces of arthropods is highly variable and the refractive index
of the \bs chitinous'' exoskeleton of insects and other classes of
animals is not universal : values from 1.52 to as much as 2 have
been reported in various studies. The melanin contents, which
increases opacity, also causes an increase of the refractive
index, a correlation which can be expected from Kramers-Kronig
causal constraints. An average index of 1.76 gives a material
which is less refractive than what we will find in the wing
substrate, so that we will refer to the overlayer material as
being \bs chitin'', emphasizing the relative lack of melanin in
this layer.

\begin{figure}[t]
\centerline{\ \includegraphics[width=7.5 cm]{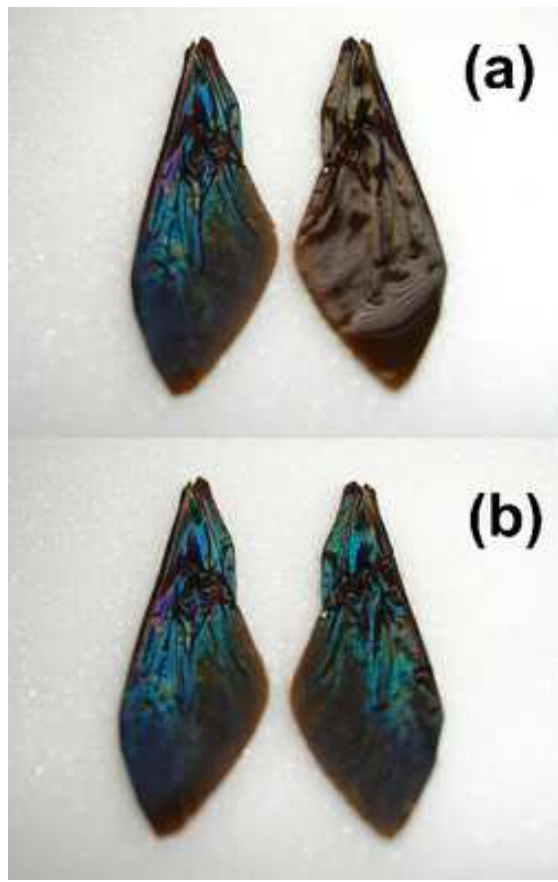}}
\caption{(Colour online) (a) In this upper panel, the wing on the
left has been kept dry as a reference, and the wing on the right
is covered with a macroscopic layer of liquid ethanol (refractive
index 1.4). The wet wing loses its blue-green iridescence and
shows a dark-brown appearance, due to the strong attenuation of
the outer interface refractive index contrast. (b) When ethanol is
removed by evaporation, the wing returns to its original
iridescent appearance.} \label{fig5}
\end{figure}

The refractive index of the opaque melanin-loaded chitin, which
makes the flexible wing substrate also calls for attention. We
first note that a single slab overlayer model implies that the
maxima of the reflection amplitude (neglecting absorption) reach
the values
\begin{equation}
\mathcal{R}_{\max }=\left( \frac{r_1+r_2}{1+r_1r_2}\right) ^2
\label{5}
\end{equation}
whereas minima of the amplitude take the values
\begin{equation}
\mathcal{R}_{\min }=\left( \frac{r_1-r_2}{1-r_1r_2}\right) ^2
\label{6}
\end{equation}
Since $r_1$ must be constant (as $n_L$ dispersion is neglected in
this model), only a strong dispersion of the substrate refractive index $n_S$ can make $%
\mathcal{R}_{\max }$ and $\mathcal{R}_{\min }$ (through $r_2$) vary with the wavelength $%
\lambda $. An experimental assessment of $\mathcal{R}_{\max }$ and $\mathcal{%
R}_{\min }$ allows to fit the real part $n_S$ of the complex index
in the following way :
\begin{equation}
n_S^{\prime }=\left\{
\begin{array}{c}
1.89\text{ if }\lambda \leq 400~\text{nm} \\
1.45\times 10^6\lambda +1.31\text{ if }\lambda \geq 400~\text{nm}
\end{array}
\right.   \label{7}
\end{equation}
An imaginary part must also be considered, as melanin involves
absorption. Following Albuquerque et al.
\cite{Albuquerque-Ebj-2006}, for visible wavelengths, this
imaginary part will be assigned the value $n_S^{\prime \prime
}\sim 0.02$. The transmission coefficient of a 5-$\mu$m thick slab
with this absorption response will not exceed 21\% at 800 nm and
will even be smaller for shorter wavelengths.

\begin{figure}[t]
\centerline{\ \includegraphics[width=7.0 cm]{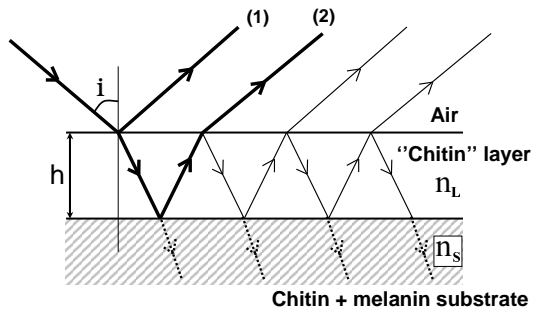}}
\caption{Schematic representation of the wing structure, showing
how the reflection takes place through the multiple paths of
light.} \label{fig6}
\end{figure}
Fig. \ref{fig4} also shows the results of calculations (black
curves) based on the above refractive indexes. This calculation
uses a simple one-dimensional coupled-modes theory, that combines
a scattering matrix formalism with a plane wave representation of
the fields. This method is well known \cite{Vigneron-spie-2006}
and do not need a detailed description. In Fig. \ref{fig4},
experimental and theoretical results are shown and compared for a
few specific angles of incidence (0$^o$, 15$^o$, 30$^o$, 45$^o$
and 60$^o$). The measured and computed location and line width of
the reflection bands correlate satisfactorily. The thin layer
model is then clearly consistent with the observed reflection
factor. The location of the maxima and minima can easily be
understood from a thin-film interference model. Indeed, given the
progression of the refractive indexes ($1$ outside, $n_L$ in the
film, $n_S$ for the substrate), constructive interference under
the incidence $i$ will occur for incident wavelengths that verify
equations (3), just as destructive interference will occur for
incidence wavelengths that verify equation(4) (see Fig.
\ref{fig6}). For instance, at normal incidence, this simple
formula predicts the following destructively reflected
wavelengths~: 671~nm, 403~nm,... These clearly agree with the
spectral observations (656~nm, 404~nm,...and match spectral
calculations.

The blue-shift of the maxima and minima of the reflection
coefficient with the increase in the angle of incidence is well
predicted. The maxima-minima damping is also well described
justifying the introduced dispersion of the melanin/chitin
mixture. In addition, calculated curves do not present significant
differences if $n_S^{\prime \prime }$ is set equal to zero (not
shown). As a consequence, melanin absorption do not play a
significant role in the present context, except for providing a
dark background which makes easier perceiving the colored
iridescence.

The color perceived from the reflection intensity can be described
by chromaticity coordinates that can be calculated in the
framework of standard human colorimetry. Fig. 7 shows the color
trajectories with the D65 illuminant, in the xy CIE 1931
chromaticity standard, as a function of the angle of incidence,
for both the theory and experimental reflectance curves.

\begin{figure}[t]
\centerline{\ \includegraphics[width=8.2 cm]{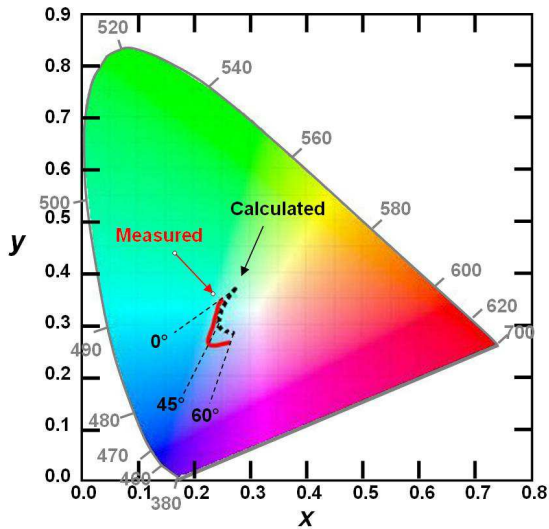}}
\caption{(Colour online) Colorimetric trajectories for light
reflected from the the wing of the wasp, for varying incidence
angles, in the range from 0 to 60$^{o}$. Solid line is from
experimental data ; dashed line is from the model calculation.}
\label{fig7}
\end{figure}

The fine details perceived in the lineshape of the $m=1$
reflection, near $\lambda $~= 1006~nm is not properly accounted by
the monolayer model. The appearance of structures on the lineshape
of this interference fringe could be due to secondary
interferences through the full thickness of the wing, but the
fundamental would appear far in the infrared, and harmonics in the
visible range should be weak, du to the strong melanin absorption.
We suggest that these structures are associated with the surface
roughness of the overlayer. As seen above, the wing surface
presents a roughness on a typical length scale of 1200~nm and, via
the decay to and from guided modes (Fano resonances)
\cite{Sarrazin-prb-2003} this can cause diffraction and addition
of spectral oscillations on the short-wavelength side of the
fringe. With a 1200~nm \bs horizontal'' grating on such structure,
diffraction can change the reflected wavelength by about $\Delta
\lambda \approx 100$~nm, consistent with the observed spectral
location of the structure (see, for instance, on Fig. 4 the weak
peak near 980~nm for $\theta=15^o$).  At the moment, however,
theory seems unable to accurately predict the observed intensities
of these fringe perturbations, so that the question  - out of the
scope of the present paper - should be given some further
attention.

\section{Conclusion}

We have shown that the iridescence of the wings of
\textit{Megascolia Procer Javanensis} can be reasonably well
understood as resulting from the interference of light in a thin
optical chitin layer covering a chitin/melanin absorbing
structure.

This wasp is equipped with opaque wings which contain a high
concentration of melanin. The black background defined by this
chitin/melanin structure allows for a particularly highly visible
structural blue-green colouration, generated by an extremely
simple device, using a minimal number of interfering waves~: a
constant-thickness overlayer covering all four wings. This is
among the most elementary interference filters and, in spite of
its simplicity, it turns out to be very effective. It is
interesting to note that, in a very different context, evolution
has produced similar structures on birds~: the domestic pigeon
\cite {Yin-pre-2006}, which displays some feathers with green
iridescence, and others with violet iridescence, use the same
strategy. In the bird, the \symbol{92} active'' overlayer is the
cortex of the barbules.

\begin{acknowledgments}
This investigation was conducted with the support of the European
NEST STREP BioPhot project, under contract no. 12915. The use of
Namur Interuniversity Scientific Computing Facility (Namur-ISCF)
is acknowledged. This work has also been partly supported by the
European Regional Development Fund (ERDF) and the Walloon Regional
Government under the "PREMIO" INTERREG IIIa project. M.R. was
supported by the Belgian Fund for Industrial and Agronomic
Research (FRIA). V.W. acknowledges the support of the Belgian
National Fund for Scientific Research (FNRS).
\end{acknowledgments}


\end{document}